\documentclass[12pt]{article}
\usepackage{amsmath,amsfonts,amssymb,latexsym}
\usepackage{graphicx}
\usepackage{color}

\newtheorem{theorem}{Theorem}[section]

\numberwithin{equation}{section}
\def\be{\begin{equation}}
\def\ee{\end{equation}}
\def\bq{\begin{eqnarray}}
\def\eq{\end{eqnarray}}
\def\beq{\begin{eqnarray*}}
\def\eeq{\end{eqnarray*}}

\def\d{\delta}

\def\t{\theta}

\begin{document}

\title{{\Huge Flat limits of curved interacting cosmic fluids}}
\author{{\Large Spiros Cotsakis\footnote{\texttt{skot@aegean.gr}}\,\,\,
 and Georgia Kittou\footnote{\texttt{gkittou@aegean.gr}}} \\
GEO.DY.SY.C. Research Group  \\ University of the Aegean \\ Karlovassi 83200, Samos, Greece}
\maketitle

\begin{abstract}
\noindent We study curved isotropic cosmologies filled with two interacting fluids near their time singularities. We find that a number of these universes asymptote to flat limits in the sense that their asymptotic properties become indistinguishable from those of flat Friedmann-Robertson-Walker models on approach to the singularity along any asymptotic direction. In particular, there are no essential singularities in these models. We discuss connections of this result with possible extensions of the cosmic no hair theorem to the case of two interacting fluids, and also provide links to a quantum cosmological treatment of real and complex Euclidean such solutions.
\end{abstract}
\section{Introduction}
Cosmic interacting fluids  have recently attracted renewed attention because of their interesting physical as well as mathematical properties. There are many astrophysical situations where an exchange of energy between two cosmic fluids becomes important, even necessary for a correct and complete description of the key cosmological processes involved. The fact that matter may possess multiple interacting components was already emphasized with the advent of the inflationary universe \cite{linde}, but also  in M-theory \cite{banks} and  in models of cosmic acceleration \cite{acc}. Such fluids admit a covariant description \cite{u},  various scaling solutions \cite{mi}, support an interesting theory of cosmological  perturbations \cite{wa} and are used abundantly in models with dark components \cite{dark}. In addition, various interesting structures in the solution space of such models \cite{chim97} as well as other, oscillator-type methods,  have been successfully  exploited to analyze more elaborate aspects of the mechanism of mutual energy transfer in specific examples such as radiation, dust and scalar fields in oscillating universes \cite{chi1}-\cite{chi4}.

In a previous paper \cite{kittou1}, we studied the dynamics near finite-time singularities of  flat isotropic universes filled with two interacting perfect fluids and found a variety  of asymptotic solutions valid locally around the spacetime singularity. Solutions include those with standard decay, phantom and cyclic universes, antidecaying models and  universes with a complicated essential singularity. In the present work, we address the question of how much the asymptotic properties of these flat solutions are altered when we consider cosmologies with curvature and keeping the interacting fluid types more or less arbitrary. In this more general case,  the interesting \emph{oscillatory} universes filled with dust or scalar fields decaying to radiation discovered by Barrow and Clifton in \cite{cl-ba07} are absent. However, we ask: Do  flat space behaviours emerge as valid and stable  asymptotic limit solutions toward a time singularity when  the effects of spacetime curvature are taken into full account? We examine all possible curved asymptotics of these models toward singularities and compare their behaviours with those of the flat space ones. When the distance between  curved and flat asymptotes becomes negligible, this may have important consequences for the flatness problem.

The plan of this paper is as follows.  In the next section, we write down all possible asymptotic decompositions of the basic system of differential equations of our problem, and emphasize the importance played by a nonlinear term present in the equations that combines the effects of curvature, expansion and fluid interaction. We call that term the \emph{exchange curvature} and discuss its significance for the dynamics of the universe induced by the curvature-fluid interaction.  Sections 3 and 4, present a detailed study of the various perturbed asymptotic solutions corresponding to power-law and phantom behaviours near spacetime singularities. In Section 5, we explain how the types of singular behaviours are altered when we pass from the flat space solutions to the curved ones asymptotically. This is useful also in the opposite direction, that is when we have a curved universe and by taking its limit to a past or a future singularity, we examine the question of whether or not  it  gets  near to a flat asymptote. In the last section we discuss our results and point out some interesting open problems in this field. Finally, in the Appendix we discuss the motivation and also various important details of the two mathematical methods used in this work, namely, the method of asymptotic splittings and the various criteria leading to a classification of spacetime singularities.

\section{Curvature and energy transfer}
We consider a curved  Friedman universe  with  scale factor $a(t)$   containing two fluids with equations of state
\begin{equation}
p_1=(\Gamma-1)\rho_1,\quad p_2=(\gamma-1)\rho_2.
\end{equation}
We take $H=\dot{a}/a $ to represent the Hubble expansion rate, and  assume a fluid interaction term of the form $s=-\beta H \rho_1+\alpha H \rho_2$, where $\alpha$ and $\beta$ are two parameters signifying the energy transfer between the two fluids. A dot means differentiation with respect to the proper time $t$. Then,  the evolution of the system is given by the Einstein equations in the form,
\begin{eqnarray}\label{evolution system}
3H^2+\frac{3k}{a^2}&=&\rho_1+\rho_2 \nonumber\\
\dot{\rho_1}+3H\Gamma\rho_1&=&-\beta H \rho_1+\alpha H \rho_2\label{arxiko} \\
\dot{\rho_2}+3H\Gamma\rho_2&=&\beta H \rho_1-\alpha H \rho_2, \nonumber
\end{eqnarray}
$k=0, \pm 1$ being the (constant) spatial curvature of the $3$-geometry. These equations admit a number of interesting exact solutions \cite{cl-ba07} showing an exchange of energy,  for instance, between radiation, dust and scalar fields. This makes universes with interacting fluids particularly appealing for further studies.

Equations (\ref{arxiko}) lead to the following master differential equation,
\be
\ddot{H}+AH\dot{H}+BH^3+\frac{k\Delta H}{a^2}=0,\label{master}
\ee
where $A=\alpha+\beta+3\gamma+3\Gamma$, $B=3(\alpha \Gamma+\beta \gamma+3\Gamma \gamma)/2$ and $\Delta=2-A+B$.
The first three terms in (\ref{master}) are well known from the flat case \cite{chim97,cl-ba06} and lead to a two-dimensional, cubic  system which has interesting dynamical characteristics studied in \cite{kittou1}. However, the last term in the master equation (\ref{master}) is new and describes the combined effects of expansion, curvature and energy exchange in this model. We therefore generally expect curvature not to  come in without some sort of energy exchange between the two fluids, however,  there are certain energy exchanges which would kill the last term while preserving a non-zero curvature, that is when energy is transferred  on the `exchange line' $\Delta =0$. This distinction will play a significant role below. Because of its special importance, we shall call the last term in Eq. (\ref{master}) the \emph{exchange curvature term}.

We shall be interested in the possible asymptotic regimes induced by the combined effect of the constant curvature and  the energy transfer between the two fluids on approach to a finite-time singularity and, in particular,  how these asymptotics  are influenced by the exchange curvature term. We may assume that the position of the time singularity is located at an arbitrary time $t_s$, but without loss of generality we take this to be zero for convenience of notation.

It will be especially useful for our considerations not to work with the master equation (\ref{master}) directly, but instead rewrite it in a suitable dynamical systems form. In this respect, we rename $H$ as $x$, introduce the new variable $z=k/a^2$, and  find  the \emph{three-dimensional}  system,
\begin{eqnarray}\label{original system}
\dot{x}&=&y\nonumber\\
\dot{y}&=&-Axy-Bx^3-\Delta xz\label{main}\\
\dot{z}&=&-2xz\nonumber,
\end{eqnarray}
 or, equivalently, the vector field $f_{(k,\Delta)}$,
\begin{equation}\label{vector field}
f_{(k,\Delta)}(x,y,z)=(y,-Axy-Bx^3-\Delta xz, -2xz).
\end{equation}
We shall study the  asymptotic dynamics of this system toward its past or future time singularities  using  the method of asymptotic splittings (see the  Appendix at the end of this work,  and also Refs. \cite{split}, \cite{1212.6737,1301.4778} for more information about asymptotics, applications and extensions).

We view the various asymptotic solutions of (\ref{original system}) as  `perturbations' of the \emph{flat space asymptotics} (that is when both $k$ and $\Delta$ are zero) studied in Ref. \cite{kittou1}. There are three kinds of such perturbations:  Those with zero exchange curvature, that is those that satisfy the equations (\ref{original system}) when $\Delta =0$; those with subdominant exchange curvature, that is  those that have a nonzero $\Delta$ term which, however, enters \emph{subdominantly} during the  evolution (cf. Appendix A.1 for this term); and  those with a \emph{dominant}, nonzero $\Delta$ term (dominant exchange curvature).

When  $\Delta=0$, we have two mutually interacting fluids in a curved universe but  the fluid-curvature interaction is switched off. This case lies in a sense  between that of having two interacting cosmic fluids in a \emph{flat} FRW universe (cf. Ref. \cite{kittou1}), and that of ultimately including the extra effects from the combined curvature-fluid interaction (either as dominant contributions or not).  Thus when $\Delta=0$, the  system (\ref{original system}) becomes
\begin{eqnarray}\label{original system 2}
\dot{x}&=&y\nonumber\\
\dot{y}&=&-Axy-Bx^3\\
\dot{z}&=&-2xz,\nonumber
\end{eqnarray}
equivalently, we have the vector field
\begin{equation}\label{vector field 2}
f_{(k,0)}(x,y,z)=(y,-Axy-Bx^3,-2xz).
\end{equation}
This  is a three-dimensional system, whereas when $k=0$, then $z=0$, and we are back to the 2-dimensional, cubic system of Ref.  \cite{kittou1}. The vector field (\ref{vector field 2}) can split (see Appendix (A.1) for more details and motivation on this) in three different ways, namely,
\begin{eqnarray}
_{I}f_{(k,0)}&=&(y,-Axy,-2xz)+(0,-Bx^3,0)\label{dec1'}\\
_{II}f_{(k,0)}&=&(y,-Bx^3,-2xz)+(0,-Axy,0)\label{decs2'}\\
_{III}f_{(k,0)}&=&(y,-Axy-Bx^3,-2xz)\label{decs3'},
\end{eqnarray}
so we may compare the possible alteration of the results with the 2-dimensional case treated in Ref. \cite{kittou1}. We note that since  solutions will have  in general at least one of their components unbounded at the finite-time singularity, we cannot simply say that because the two systems reduce to one another in the case of zero curvature, the same is true for their \emph{solutions}. Therefore a full asymptotic analysis is necessary presently.

When $\Delta\neq 0$, the vector field (\ref{vector field}) splits in seven different ways of the form `dominant part plus subdominant part',
 \be
 _{j}f_{(k,\Delta)}=_{j}f^{(0)}_{(k,\Delta)}+_{j}f^{(\textrm{sub})}_{(k,\Delta)},\quad j=I,\cdots,VII,
 \ee
 divided into two main groups according to whether or not the last term in the second equation in  the system (\ref{main}) is dominant or subdominant asymptotically. There are three decompositions having the $\Delta$-term subdominant, namely,

 \emph{Group A: $\Delta$-term subdominant}
 \begin{eqnarray}\label{decs1}
_{I}f_{(k,\Delta)}&=&(y,-Axy, -2xz)+(0,-Bx^3- \Delta xz,0)\\
_{II}f_{(k,\Delta)}&=&(y,-Bx^3, -2xz)+(0,-Axy- \Delta xz,0) \label{decs2}\\
_{III}f_{(k,\Delta)}&=&(y,-Axy-Bx^3, -2xz)+(0,- \Delta xz,0), \label{decs3}
\end{eqnarray}
and another four containing that  term as an additive contribution in their dominant components, namely,

 \emph{Group B: $\Delta$-term dominant}
\begin{eqnarray}
{}_{IV}f_{(k,\Delta)}&=&(y,-\Delta xz, -2xz)+(0,-Axy-Bx^3,0)\label{decs4}\\
{}_{V}f_{(k,\Delta)}&=&(y,-Axy -\Delta xz,-2xz)+(0,-Bx^3,0)\label{decs5}\\
{}_{VI}f_{(k,\Delta)}&=&(y,-Bx^3 - \Delta xz, -2xz)+(0,-Axy,0)\label{decs6}\\
{}_{VII}f_{(k,\Delta)}&=&(y,-Axy-Bx^3 - \Delta xz, -2xz)\label{decs7}.
\end{eqnarray}
Here, the first part in each decomposition, $_{j}f^{(0)}_{(k,\Delta)}$, represents the dominant collection of terms in  the given splitting, while the second one represents all subdominant terms, $_{j}f^{(\textrm{sub})}_{(k,\Delta)}$.

Below we look for Fuchsian-type solutions  valid asymptotically near the singularity at $t=0$, that is formal series expansions of the general form,
\be
X(t)=t^{h}\sum_{i=0}^{\infty }c_{i}t^{i/s},
\ee
with $X=(x,y,z)$, in particular, having no constant term\footnote{Formal series starting with a constant term and having rational exponents are called \emph{Puiseux series} and are connected with the existence of sudden singularities. It is interesting that the `flat system', that is (\ref{original system}) with $k=0, z=0$, does not admit any sudden singularities.}.  The first term in these  expansions is a scale-invariant, exact solution of the asymptotic system (see Appendix (A.1) for background on this),
\be
\dot X(t)=_{j}f^{(0)}_{(k,\Delta)},
\ee
of the form (this is the \emph{dominant balance})
\be
\mathcal{B}=[(\theta,\xi,\nu),(p,q,r)],
\ee
where
\be
x(t)=\theta t^p,\quad y(t)=\xi t^q,\quad z(t)=\nu t^r.
\ee
Here, the dominant coefficients $c_0=(\theta,\xi, \nu)\in\mathbb{C}^3$, the dominant exponents $h=(p,q,r)\in\mathbb{Q}^3$, the denominator in the indicial exponents $s\in\mathbb{Z}$, and the remaining expansion coefficients $c_i=(c_{1i},c_{2i},c_{3i})$ in the series developments of the three unknowns $x, y, z$ are all to be determined by the asymptotic analysis. We note that for any value of the curvature, $k=0$, or $\pm 1$, and each possible dominant balance, one needs to check the following  \emph{condition of subdominance} (see Appendix (A.1) for an explanation of this condition),
\begin{equation}\label{sublimit}
\lim_{t\rightarrow 0}\frac{f^{\,\textrm{sub}}_{(k,\Delta)}(c_0\,t^{h})}{t^{h-1}}= 0,
\end{equation}
for the corresponding decomposition to be admissible \cite{split}. The full solutions constructed recursively in this way, will  have valid asymptotic expansions in a neighborhood of the time singularity.

\section{Perturbing the flat attractor}
The decomposition  (\ref{dec1'}) represents the curved generalization of the flat space splitting $_{I}f_{(0,0)}=(y,-Axy)$, which as we know admits a general asymptotic solution of the form \cite{kittou1},
\be \label{expb1}
x=\frac{2}{A}t^{-1}+c_{21}t-\frac{A}{10}c_{21}^2t^3\cdots ,
\ee
that is we have the leading order form
\be\label{sol1}
H\sim \frac{2}{A}t^{-1},\quad\textrm{or}\quad a(t)\sim t^{2/A}.
\ee
This is an attractor of all smoothly evolving solutions at early  times valid \emph{for all values of} $A$ in the flat case: As shown in Ref.  \cite{kittou1}, it attracts all solutions in the Barrow-Clifton family,
\be\label{ba-cl1sol2}
a_{\textrm{BC}}(t)\sim\left(t^{-2/|A|}\right)^{1/(1-8\d)^{1/2}},\quad\d\in
[0,1/8),
\ee
in the asymptotic sense that
\be
a_{\textrm{BC}}(t)\rightarrow t^{-2/|A|},\quad\textrm{as}\quad
\d\rightarrow 0.
\ee
When we pass on to the curved case of the first decomposition of the system (\ref{dec1'}) (this is the vanishing exchange curvature case), the required subdomimant condition (\ref{sublimit}) is satisfied \emph{only if} $B=0$. Since $\Delta=2-A+B=0$, we arrive at the conclusion that the parameter $A$ must necessarily take the value
\be
A=2,
\ee
in this case for the whole scheme to be acceptable. The balances  then read,
\begin{eqnarray}
{}_{I}\mathcal{B}_{1}&=&[(1,-1,c),(-1,-2,-2)], \\
{}_{I}\mathcal{B}_{2}&=&[(1, -1,0),(-1,-2,r)]\label{2nd bal}.
\end{eqnarray}
The $\mathcal{K-}$exponents of these balances are given by the forms (see Appendix (A.1) for an explanation of this),
\begin{eqnarray}
\textrm{spec}(_{I}\mathcal{K}_1)&=&(-1,0,2),\label{IK1'}\\
\textrm{spec}(_{I}\mathcal{K}_2)&=&(-1,2,-2-r),\label{IK2'}
\end{eqnarray}
where $r$ is an arbitrary constant (below we choose it such that we have the corresponding eigenvalue positive, otherwise the final solutions will have less arbitrary constants than those required for them to be general ones).
 For the  balance $_{I}\mathcal{B}_{1}$, the  asymptotic solution approaching the singularity is then  given by,
\begin{eqnarray}
x(t)&=&t^{-1}+c_{21}t+\cdots\label{sol10},\\
z(t)&=&ct^{-2}+3t^{-1}-cc_{21}+\cdots .\nonumber
\end{eqnarray}
The compatibility condition at the $j=2$ level with the associated eigenvector $v^{T}=(1,1,-c)$ reads
\begin{equation}
c_{23}=-cc_{21},
\end{equation}
and this is found to be identically satisfied after recursive calculations. Therefore, since it contains three arbitrary constants, that is $c,c_{21}$, and the position of the time singularity, (\ref{sol10}) is a general solution of the cosmological equations (\ref{original system 2}) valid in the neighborhood of the finite-time singularity.
We therefore conclude that the curved perturbation of the flat attractor when the exchange curvature vanishes restricts the possible solutions  and allows only a collapse type singularity $(A=2>0)$, and no big-rip types like in the flat case\footnote{The asymptotic series solution constructed using  the second balance, (\ref{2nd bal}), of the decomposition  is similar to the solution (\ref{sol10}) for $x$, but  the series expansion for $z$ equals to zero. Hence, we disregard this as an acceptable curved solution.}.  In contrast to the latter case where we had the full, one-parameter ($A$)-family of solutions, cf. \cite{kittou1}, here there can be no attractor property in the sense we met in the flat case, except for that specific $A$-value.

Let us now move on to include a nonzero $\Delta$ term in the perturbation (\ref{dec1'}), that is consider it when the exchange curvature is turned on. There are two ways for such a term to be included in (\ref{dec1'}), first in the subdominant part of the splitting as in decomposition (\ref{decs1}), and secondly in the dominant contributions, cf. (\ref{decs5})\footnote{Thus, exchange curvature may be regarded as  a truly dynamical form of `curvature' for it can play a different role during the evolution in the sense that it describes the interaction between the constant curvature $k$, the fluids' exchange of energy given by the parameter $\Delta$, and the expansion $H$ in these models.}.

For (\ref{decs1}), the first thing to notice is that since the change occurs only in the subdominant part and the dominant system is identical to the one from (\ref{dec1'}), the asymptotic dynamics of the universe in this case will have the same scale invariant solutions, namely, the dominant balances
\begin{eqnarray}
\label{1.1}{}_{I}\mathcal{B}_{1}&=&[(2/A, -2/A,c),(-1,-2,-4/A)] \\
\label{1.2}{}_{I}\mathcal{B}_{2}&=&[(2/A, -2/A,0),(-1,-2,r)].
\end{eqnarray}
However, the candidate \emph{subdominant} part is now different, $_{I}f^{\,\textrm{sub}}_{(k,\Delta)}=(0,-Bx^3-\Delta xz,0)$, and instead of giving the previous restrictions on $A,B$, it becomes asymptotically subdominant only if we set
\be\label{range b1}
B=0,\quad \textrm{and}\quad 2-4/A>0,
 \ee
for the first balance (\ref{1.1}), whereas for the second balance (\ref{1.2}), the subdominant condition is satisfied so long as $B=0$. Further, the spectra of the corresponding Kovalevskaya matrices  are given by,
\begin{eqnarray}
\textrm{spec}(_{I}\mathcal{K}_1)&=&(-1,0,2)\label{IK1},\\
\textrm{spec}(_{I}\mathcal{K}_2)&=&(-1,2,-4/A+2)\label{IK2},
 \end{eqnarray}
respectively. The first balance, (\ref{1.1}), then leads to an acceptable curved solution in the form of a formal expansion \emph{only} in the range $A<0$ (of the two possible cases from (\ref{range b1})), the final result is,
\begin{eqnarray}\label{expb3}
x(t)&=&-\frac{1}{2}t^{-1}+c_{21}t+\cdots\label{sol1},\\
z(t)&=&ct^{-1}-cc_{21}+\cdots ,\nonumber
\end{eqnarray} while the $y$ series expansion is obtained from the $x$ series by differentiation.
The expansion (\ref{sol1})  is indeed a general solution. Here, for concreteness we have given it for the choice $A=-4$. We note that the second balance,  (\ref{1.2}), does not lead to curved solutions and  so we do not discuss it further.

We therefore conclude that curved perturbations of the flat solution with the $\Delta$ term nonzero but subdominant asymptotically, Eq.(\ref{sol1}), exist in the range $A<0$, are similar to the $\Delta=0$ curved perturbations of the same solution (cf. (\ref{sol10}), but have  somewhat milder behaviour for $z$.

The last kind of possible curved perturbations of the flat attractor (\ref{dec1'}) involves  a nonzero exchange curvature  term which, however, now appears in the \emph{dominant} contributions asymptotically. This is precisely the scheme given in the asymptotic splitting  (\ref{decs5}). The candidate subdominant term ${}_{V}f^{(\textrm{sub})}=(0,-Bx^3,0)$ is asymptotically subdominant only if $B=0$. The $\mathcal{K-}$exponents are found to be
\begin{eqnarray}
\textrm{spec}(_{V}\mathcal{K}_1)&=&(-1,2,2-A),\label{VK1}\\
\textrm{spec}(_{V}\mathcal{K}_2)&=&(-1,2,2-4/A),\label{VK2}
\end{eqnarray}
and we see that, again,  depending on the sign of parameter $A$, the number of positive eigenvalues varies. Let us take the first spectrum (\ref{VK1}) and $A\leq 2$. Then the asymptotic solution is expected to be general if arbitrary coefficients appear at the $j=1$ and $j=2$ levels in the series expansions for the specific choice $A=1$. The final solution is given by
\begin{eqnarray}
x&=&t^{-1}+c_{11}+c_{21}t+\cdots,\label{sol9}\\
z&=&-t^{-2}+2c_{11}t^{-1}+c_{21}+\cdots,\nonumber
\end{eqnarray}
and indeed represents a general series expansion.
The compatibility conditions at the $j=1$ and $j=2$ levels (with  associated eigenvectors  $v^{T}=(1,0,2)$ and $v^{T}=(1,1,1)$) are given by
\begin{eqnarray}
c_{11}+c_{12}-\frac{1}{2}c_{13}&=&0\\
3c_{21}+(1-A)c_{22}-(4-A)c_{23}&=&0,
\end{eqnarray} and are true after recursive calculations.
For $A>2$,  only one positive $\mathcal{K-}$exponent exists and the solution becomes a particular one in this case. Because of the special form of the leading order for $z$ in this solution, we conclude that (\ref{sol9}) applies only to open models.

Finally,  the second balance in the parameter range $0<A<2$ does not lead to any nontrivial curved solutions and we do not discuss it further, whereas when  $A<0$ the  solution is analogous to (\ref{sol1}) thus leading to a big rip for both open (when $c<0$) and closed (when $c>0$) universes. For  $A\geq2$, the general solution is given by
\begin{eqnarray}
x&=&\frac{2}{A}t^{-1}+c_{11}+c_{21}t+\cdots,\label{sol2}\\
z&=&c_{13}t^{-1}+c_{23}+\cdots.\nonumber
\end{eqnarray}

\section{Phantom perturbations}
In the section, we consider the curved generalizations of the flat space splitting $_{II}f_{(0,0)}=(y,-Bx^3)$. In particular, we perturb the flat space solutions which lead to the  phantom big bang ($+$ signed) or big rip ($-$ signed, see below) singularities for $B<0$, cf. \cite{kittou1}, Section 4\footnote{We had given only the collapse type in that reference.}. This is the only  surviving case presently as we show in this section. We recall that  in the flat case, the general asymptotic solution is of the form,
\be \label{expb2}
x=\pm\sqrt{-2/B}\,t^{-1}+c_{41}t^3\mp\frac{B}{12}c_{41}^2\sqrt{-2/B}\,t^7+\cdots,
\ee
so to leading order we have,
\be\label{sol3}
H\sim \pm \sqrt{-2/B}\,t^{-1},\textrm{or}\quad a(t)\sim t^{\pm\sqrt{-2/B}},
\ee
and this is an asymptotic attractor of all smoothly evolving phantom solutions on approach to the singularity \cite{kittou1}.

When we pass on to the curved case, first from the splitting (\ref{decs2'}) corresponding to $\Delta=0$ (the vanishing exchange curvature case), and also from the decomposition (\ref{decs2}) that is when the exchange curvature enters subdominantly, we find that  the asymptotic series for the $z$ expansion is zero identically, and so  the overall dynamics stemming from these two cases  is of no interest as they  do not represent a curved solution.

We then focus on the asymptotic behaviour of curved perturbations of the flat attractor involving the exchange curvature term in the \emph{dominant} contributions asymptotically. This is precisely the scheme given in the asymptotic splitting  (\ref{decs6}). The candidate subdominant term $_{VI}f^{(\textrm{sub})}=(0,-Axy,0)$ is asymptotically subdominant only if $A=0$. The $\mathcal{K-}$exponents are given by the forms
\begin{eqnarray}
\textrm{spec}(_{VI}\mathcal{K}_1)&=&(-1,4,2-2\sqrt{-2/B}),\label{VIK1}\\
\textrm{spec}(_{VI}\mathcal{K}_2)&=&(-1,4,2+\sqrt{-2/B}).\label{VIK2}
\end{eqnarray}
The only interesting range of values is with the first spectrum (\ref{VIK1}) and $B\leq -2$. Then the asymptotic solution is expected to be general if arbitrary coefficients appear at the $j=1$ and $j=4$ level in the series expansion for the specific choice $B=-8$. So the solution is given by
\begin{eqnarray}
x(t)&=&\frac{1}{2}t^{-1}+c_{11}-c_{11}(1-c_{13})t+\cdots\label{sol4},\\
z(t)&=&-2c_{11}t^{-1}-4c_{11}+\cdots \nonumber
\end{eqnarray}
and indeed represents a general series expansion since the compatibility conditions at the $j=1$ and $j=2$ level (with  associated eigenvectors  $v^{T}=(1,0,2)$ and $v^{T}=(1,1,1)$)  are found true after recursive calculations. It can be seen that this asymptotic behaviour  leads to another big bang singularity in this case.

The second balance (\ref{VIK2}) gives general results only in the range $B<0$, and there it  leads to an asymptotic  big rip  solution.
 Choosing the  value $B=-8$, it reads,
\begin{equation}
\label{2.1}{}_{II}\mathcal{B}_{1}=[(-1/2,1/2,0),(-1,-2,-2)].
\end{equation}
The candidate subdominant part, $_{II}f^{\,\textrm{sub}}_{(k,\Delta)}=(0,-Axy-\Delta xz,0)$ becomes truly asymptotically subdominant only if we set $A=0$.
The spectrum of the corresponding Kovalevskaya matrix is given by
\begin{equation}
\textrm{spec}(_{II}\mathcal{K}_1)=(-1,3,4)\label{IIK1},
 \end{equation}
and the final formal expansion takes the form
\begin{eqnarray}
x(t)&=&-\frac{1}{2}t^{-1}+c_{11}-c_{11}(1-c_{13})t+\cdots\label{sol3a},\\
z(t)&=&2c_{11}t^{-1}-4c_{11}+\cdots. \nonumber
\end{eqnarray}
The compatibility conditions at the $j=3$ and $j=4$ levels (with  associated eigenvectors  $v^{T}=(3,6,4)$ and $v^{T}=(1,3,0)$) are  found to be identically satisfied  after recursive calculations. Thus (\ref{sol3a}) contains three arbitrary constants, $c_{11}, c_{13}$, and the position of the singularity (which we have taken it to be at $t=0$ without loss of generality).

\section{Extremality types}
How does the \emph{type} of singularity change (or does not change) as we pass from the flat to the curved solutions considered in the previous sections? As discussed in detail in the Appendix (A.2),  cosmological singularities may be  classified into triplets of the form $(\textrm{SF}_i, \textrm{BR}_j, \textrm{H}_k)$:
There are three broad singularity types, $\textrm{SF}_{i}, i=1,2,3$, according to the leading behaviour of the scale factor: Collapse ($a\rightarrow 0$), sudden ($a\rightarrow a_s\neq 0$), and big rip ($a\rightarrow\infty$) singularities  respectively.
Each $\textrm{SF}_i$ type  is subdivided into four $\textrm{BR}_{j}, j=1,2,3,4$ categories of singularities formed according to the behaviour of the \emph{Bel-Robinson energies} which signify the \emph{material} character of the singularities. For the  isotropic universes we consider here, these energies have only their \emph{electric} parts nonzero, namely, the functions $|E|^2=3(\ddot{a}/{a})^2$ and $|D|^2=3((\dot{a}/{a})^2+{k}/a^2))^2$.  $|D|$ and $|E|$ are proportional to the total fluid density $\rho_{tot}$ and pressure $p_{tot}$ respectively. Thirdly, we need to take into account  the behaviour of the spatial slices asymptotically. This is monitored by the possible anomalies in the evolution of the Hubble expansion rate, $H$, and gives the remaining, \emph{geometric}, part of the character of the singularities. It can be shown (cf. Appendix (A.2)) that the possible erratic behaviour of $H$ is exhausted if we introduce three further  classes, $\textrm{H}_{k}, k=1,2,3$, according to which either $H$ is not piecewise continuous, or it blows up in a finite time, or, finally,  it is integrable only for a finite time duration.

For the flat space family of asymptotic solutions (\ref{expb1}), we find that the structure of their singularities depends on the sign of the parameter $A$, and it is of the type
\begin{itemize}
\item $(\textrm{SF}_1, \textrm{BR}_2, \textrm{H}_2)$, when $A>0$, and
\item $(\textrm{SF}_3, \textrm{BR}_2, \textrm{H}_2)$, when $A<0$.
\end{itemize}
In both cases, the total energy density and the total pressure diverge as we approach the singularity (type $\textrm{BR}_2$). To construct a specific physical example, suppose we have the  vacuum decay model of \cite{cl-ba06} representing the decay of a vacuum stress into  radiation. When $A>0$,  using the subdominance condition we have  $B=0$,  and the energy exchange parameters $\alpha$ and $\beta$ ($A=\alpha+\beta+3\Gamma+3\gamma$, $B=3/2(\alpha\Gamma+\beta\gamma+3\Gamma\gamma)$) satisfy:
$
-4<\alpha<\infty,\, \beta=0.
$
The dominant part of the asymptotic series solution for the scale factor is  $a(t)\sim t^{2/A}$, and for $A>0$, it collapses to zero size $\alpha\rightarrow  0$ on approach to the singularity, giving an $\textrm{SF}_1$ type. The  electric parts of the Bel-Robinson energy (see Appendix) are given by the forms,
\be
 |E|^2=12\left(\frac{2-A}{A^2t^2}\right)^2,\quad |D|^2=3\left(\frac{4}{A^2t^2}\right)^2,
\ee
and so they  are asymptotically diverging $(p_{\textrm{tot}}\rightarrow \infty, \rho_{\textrm{tot}}\rightarrow \infty)$ leading to a  $\textrm{BR}_2$ type. Last, since  $H\sim (2/A)t^{-1}$, we find that the singularity type  is also of an  $\textrm{H}_2$ type. When $A<0$, $B=0$, and the energy exchange parameters satisfy,
$
-\infty<\alpha<-4,\, \beta=0.
$
Asymptotically the dominant part of the scale factor leads to a big rip ($a\rightarrow\infty$), and so the singularity is an $\textrm{SF}_3$. The total pressure and total energy density have the same behaviour as in case $A>0$, therefore the type of singularity in this example  is also $\textrm{BR}_2$, whereas the Hubble parameter is given by $H\sim (2/A)t^{-1}$ with $A<0$.  The type of singularity for the Hubble parameter that describes the divergence of $H$ at future finite time is $\textrm{H}_2$.

The same structures appear for the phantom solutions (\ref{expb2}) in the allowed $B$ range (we have taken $B=-8$), the singularity is of the type $(\textrm{SF}_1, \textrm{BR}_2, \textrm{H}_2)$ for the $(+)$ branch, and of the second type, $(\textrm{SF}_3, \textrm{BR}_2, \textrm{H}_2)$,  for the $(-)$ branch.   To construct a specific example, let us consider the exchange of energy between a ghost fluid with $\Gamma<-1$ and radiation (with $\gamma=4/3$). Then  the $(+)$ branch of the solution has
$
-\infty<\alpha<2,\, 1<\beta<\infty,
$
and the dominant part of the scale factor satisfies $a(t)\sim t^{1/2}$  and so tends to zero  (type $\textrm{SF}_1$). Then the forms of $|E|^2$ and $|D|^2$, become $|E|^2=|D|^2=(3/16)t^{-4}$, therefore both total pressure and total energy density diverge for $t\rightarrow 0$ (type $\textrm{BR}_2$), whereas for the Hubble parameter we have the singularity type $\textrm{H}_2$. Gathering all the information about the singularity type for the $(+)$ branch of solution (\ref{expb2}) we conclude that is of said type. As for the $(-)$ branch of the solution, we find that the energy exchange parameters satisfy the same range of values as in the $(+)$ branch.  The  scale factor asymptotically blows up as $a(t)\sim t^{-1/2}$, while the total pressure and total energy density asymptotically  diverge since $|E|^2=(27/16)t^{-4}$ and $|D|^2=(3/4)t^{-2}$--type $\textrm{BR}_2$. The dominant form of Hubble parameter $x=H\sim -(1/2)t^{-1}$--type   $\textrm{H}_2$.

When we pass on to the curved models with zero exchange curvature, we have only the asymptotic power-law solutions (\ref{sol10}), and these are possible only for the positive value of $A=2$. In particular, there are no phantom solutions with zero exchange curvature. Then, we find that the collapse singularity of the flat geometry (in the $A>0$ case) survives in this case, but its material character changes into an:
\begin{itemize}
\item $(\textrm{SF}_1, \textrm{BR}_4, \textrm{H}_2)$, when $k=+1$, and an
\item $(\textrm{SF}_1, \textrm{BR}_3, \textrm{H}_2)$, when $k=-1$,
\end{itemize}
while its geometric character remains the same. Thus either the density and pressure become negligible asymptotically (open models with a $\textrm{BR}_3$ type), or we only have the density diverging while the pressure vanishes asymptotically. That is we find closed universes  with a type $\textrm{BR}_4$ big bang. To our knowledge,  such a singular behaviour appears here for the first time. To construct specific models with such behaviours we can take again the decaying vacuum example considered also above. Then for an open universe and $A=2, B=0$, the energy exchange parameters are given by
$
\alpha=-2,\,\beta=0.
$
The dominant part of the scale factor is given by $a(t)\sim t$ and asymptotically collapses to zero size ($\textrm{SF}_1$ type), and so we find that $|E|^2\rightarrow0$ and $|D|^2\rightarrow0$, leading to the singularity type $\textrm{BR}_3$. In addition, the dominant form of Hubble parameter is $H\sim t^{-1}$, and so blows up at $t=0$.  Consequently  the type of singularity is $\textrm{H}_2$. So for open decaying vacua, the singularity type for solution (\ref{sol10}) is of the type
$(\textrm{SF}_1, \textrm{BR}_3, \textrm{H}_2)$.
When $k=+1$, the energy exchange parameters are the same as above. The geometric character of the singularity remains the same as in the case of open universes (scale factor collapses to zero size $(\textrm{SF}_1)$, Hubble parameter blows up $(H_2)$).   The material behaviour of the singularity, however, changes since the total pressure asymptotically becomes negligible whereas the total energy density diverges. That is to say such closed universes  have a singularity type of the class $\textrm{BR}_4$.

Let us finally comment on the case of the asymptotic solutions with a \emph{nonzero} exchange curvature term. This includes the singular big rip solutions having a  nonzero exchange curvature term in their \emph{subdominant} parts asymptotically, as in power-law solution  (\ref{sol1}). This will give behaviours of the type $(\textrm{SF}_3, \textrm{BR}_2, \textrm{H}_2)$ for both values of the curvature $k$. The vacuum-radiation exchange model for different parameter values is a suitable family for examples in this case too. If we take $A<0$, $B=0$, then we  have
$
-\infty<\alpha<-4,\,\beta=0.
$
Then  $a(t)\sim t^{-1/2}$  diverging at the singularity--type $\textrm{SF}_3$. For both closed $(k=+1)$ and open $(k=-1)$ universes, the electric parts  of the Bel-Robinson energy are given by $|E|^2=27/16t^{4}$ and $|D|^2=3(1\pm_{(k=\pm 1)} 4t)^2/(4t^2)^2$. Hence,  the type of singularity is $\textrm{BR}_2$. Based on the dominant part of the asymptotic series solution, the Hubble parameter $H\sim -1/2t$ possesses a singularity of type $H_2$.

When $\Delta$ is dominant asymptotically, we have the  singular asymptotics (\ref{sol9}), (\ref{sol2}) and the phantom solutions  (\ref{sol4}), (\ref{sol3a}). The singularity for the open universe solution (\ref{sol9}) is similar to that found in the case with a zero $\Delta$. To see an example of this, we take $A=1$, $B=0$ and the energy exchange parameters become
$
\alpha=-3,\,\beta=0.
$
This universe emerges from a big bang at $t=0$ with $a(t)\sim t\rightarrow0$. The total pressure and  energy density for the model in question become trivial asymptotically and the  Hubble parameter diverges at $t=0$. So we have a singularity of type $ (\textrm{SF}_1, \textrm{BR}_3, \textrm{H}_2)$ in this case.

When $k=+1$, the singularities for (\ref{sol2}), (\ref{sol4}) are of the form $(\textrm{SF}_1, \textrm{BR}_2, \textrm{H}_2)$, i.e., big bangs with both the pressure and the energy density asymptotically diverging.
For (\ref{sol2}), such a  singularity can be described when we choose $A\geq2$, $B=0$, $\Gamma=0$ and $\gamma=4/3$, so that
$
-2\leq \alpha<\infty,\,\beta=0.
$
A specific  example for the singularity in the solution (\ref{sol4})  for $k=+1$ can be obtained by taking the interacting fluids to be  a ghost fluid with $\Gamma<-1$ and choose the second one to be radiation, $\gamma=4/3$. Then the parameters $A$ and $B$ satisfy $A=0$ and $B=-8$, and so we are led to the choice
$
\alpha=-4/3,\,\beta=-8/3.
$
One may easily see that all conditions for such a behaviour are satisfied.

There are also big rip singularities coming from the forms (\ref{VK2}), (\ref{VIK2}) and (\ref{sol3a}) for both closed and open models. For  $A<0$ and $B=0$, the form (\ref{VK2}) has a solution analogous to (\ref{expb3}), whereas an example for the behaviours (\ref{VIK2}) and  (\ref{sol3a}) is to take $\Gamma<-1$, $\gamma=4/3$, with the energy exchange parameters  given by
$
\alpha=0,\,\beta>-1.
$

\section{Discussion}
In this paper, we analyzed the asymptotic stability of the singular, flat space solutions with two interacting fluids against isotropic, curved  perturbations on approach to a finite-time singularity. We have found that only two of the flat space solutions survive, namely the flat space power-law attractor and the phantom solutions. We have examined the structure of the resulting perturbations and found a detailed array of possible behaviours for closed and open universes, five inequivalent power-law asymptotics and two phantom solutions in all.

For specific choices of the fluids and their interactions, these curved asymptotics become indistinguishable from their flat counterparts as it may be seen directly from their specific forms. This may have important consequences for the resolution of the flatness problem in our asymptotic context of two interacting cosmic fluids. In addition, our results set a firm basis for stability questions in the more general context of two fluids that exchange energy. When the two fluids are ultra-stiff and satisfy $\gamma>\Gamma>2$, the curved solutions of Section 3 approach the corresponding flat space asymptote. This is very important, for it provides a first step of generalizing the known, single-fluid, stability result at ultra-stiff singularities of Refs. \cite{eric}, \cite{lid} to the case of two interacting fluids with the type of interaction considered here. It is an open question whether the ultra-stiff stability of two \emph{interacting} cosmic fluids toward a past singularity proved here extends to the case of anisotropic perturbations, thus giving a cosmic no hair result for cosmologies with two interacting fluids.

Another basic property of the resulting solutions is that all singularities are first- or second-order poles, in particular, there are no essential singularities in these models like in the flat case. All asymptotic solutions lead to either a big bang or a big rip singularity and there are no sudden (weak, in other terminology) singularities in isotropic universes with interacting fluids with the particular kind of interaction considered in these limits.

Let us briefly comment on the impossibility of constructing stable perturbations of flat asymptotics of types other than those considered in this work. We first examine  curved perturbations of the flat \emph{anti-decaying and cyclic universes} found in \cite{kittou1}.
Such universes were constructed as stable, flat models asymptotically on approach to the singularity from the flat decomposition $f^{(3)}=(y,-Bx^3)$, in the notation of \cite{kittou1}. Do such behaviours remain valid when we perturb them in an analogous way as above? In other words, do, say, antidecaying curved  universes become flatter asymptotically? Any  curved perturbation of such models with $\Delta=0$ would arise from the
 second decomposition of the system, namely, the form (\ref{decs2'}), while their curved perturbations with a nonzero  $\Delta$ term would lie in the asymptotic splittings (\ref{decs2}) when this term is subdominant asymptotically for $B>0$, and (\ref{decs6}) when it is dominant also for $B>0$.
 We have performed a detailed analysis  of all the possible asymptotics in such cosmologies and we find that any such perturbation cannot be sustained asymptotically as a valid solution of the curved field equations (\ref{original system}). These calculations will be included elsewhere  \cite{kittouPhD}. A similar negative conclusion applies to the curved perturbations of the various flat \emph{decaying
  asymptotics}, i.e., those stemming from the flat decomposition  $f^{(1)}=(y,-Axy-Bx^3)$ constructed in Ref. \cite{kittou1}. Such models would follow from (\ref{decs3'}) when $\Delta=0$, and from (\ref{decs4}) and (\ref{decs7}) when $\Delta\neq 0$. All these asymptotics cannot exist consistently as solutions of the basic system  (\ref{original system}).

It is in principle possible that a more general type of fluid interaction will alter the character of the singularities and asymptotics  analyzed in this work. It may be that other types of behaviour, not appearing here, will emerge and become possible with a more general type of fluid interaction. This is a problem currently under study.

In this work we have focused on \emph{real} solutions, in particular, we have taken the flat and curved universes filled with interacting fluids asymptotically toward the finite-time singularity \emph{along the real time direction} and proved that all solutions are pole-like toward the singularity. However, the method of asymptotic splittings is genuinely one that builds \emph{complex} dynamical solutions, the unknowns will in general be complex functions of a complex (time) variable. Therefore one may naturally wonder how a rotation to imaginary time of the real solutions will affect the resulting Euclidean solutions, that is will there be any singularities along the imaginary direction? In the present work, we have shown that all singularities  are poles, and so they will remain poles when $t$ approaches the singularity at $t=0$ \emph{from any direction} in the complex plane. The only hope is then when we have an \emph{essential} singularity somewhere, in the sense that only in that case there is a chance of an asymptotic direction where the Euclidean  model will be free of singularities. This is indeed the case of the flat universe with an essential singularity found in \cite{kittou1} (cf. solution (6.12) in that reference). It would be exceedingly interesting to have a quantum cosmological argument to distinguish when the wave functions of these real or complex, nonsingular Euclidean and singular Lorentzian classical solutions with interacting fluids will be oscillatory or convergent.

\section*{Acknowledgements}
We are grateful to an anonymous referee whose comments and suggestions have led to an improved version of this paper. The work of G.K. was supported by a PhD grant co-funded by the European Union (European Social Fund-ESF) and national resources under the framework `Herakleitus II: Action for the enforcement of human research potential through the realization of doctorate work' which is gratefully acknowledged.

\appendix
\section{Appendix: Asymptotic splittings and singularities}
In this Appendix, we collect together for easy reference of the reader a number of results concerning the two main mathematical methods used in the main body of this work, namely, asymptotic splittings and criteria for the classification of singularities. We discuss the physical motivation and explain various important issues about the use of these methods and their relevance to the physical problem addressed here.

\subsection{Method of asymptotic splittings}
Suppose that, motivated by the problem (\ref{original system}), (\ref{vector field}), we are interested in the dynamics and nature of the solutions near a finite-time singularity of the general dynamical system
\be\label{general}
\dot x=f(x),\quad x=x(t)\in\mathbb{R}^n,
\ee
defined by the vector field $f$. We are interested in particular in the nature of  solutions of the dynamical system (\ref{general}) near a time $t_*$ (which may be taken to be zero, $t_*=0$) such that the following property is satisfied,
\begin{equation}
\lim_{t\rightarrow t_{*}}|{x}(t;{x}_{0})|=\infty ,  \label{sing2}
\end{equation}%
which means that at least some component of the $n-$dimensional solution vector $x(t)$, developed from the initial data $x_0$, blows up at the finite-time singularity located at $t_*$.

We then suppose that on approach to the finite time singularity the vector field $f$ shows some dominant feature (otherwise the field whirls around the singularity in a spiral, bounded way without showing any blow up) and therefore decomposes (or splits) into some dominant part ${f}^{\,(0)}$ and another, subdominant part, as follows:
\begin{equation}
{f}={f}^{\,(0)}+f^{(\textrm{sub})}.
\label{dec1}
\end{equation}
In particular, we look for weight-homogeneous decompositions where $f^{(\textrm{sub})}$ has a nonzero vector weighted degree (cf. Refs. \cite{split,1212.6737,1301.4778}). In both the dominant and  subdominant  parts, one has to try all possible combinations of terms in  the vector field  $f$ as candidates. For instance, the 2-dimensional vector field $f=(y,x^2+y)$ can in principle split as above in three possible ways, with  `dominant' parts  $(y,x^2),(y,y)$, or $(y,x^2+y)$, and `subdominant' parts $(0,y),(0,x^2),(0,0)$ respectively (the third splitting corresponds to the so-called all-terms-dominant case).

We thus see that the asymptotic decomposition of $f$ is highly nonunique, however, whether or not any of these candidate splittings will be finally acceptable is something to be determined by the subsequent asymptotic analysis (cf. \cite{split} and also below). In particular, there are two  important asymptotic conditions to be satisfied in order to determine that a given  splitting is dynamically acceptable. The first concerns the dominant part of the vector field $f$, in particular, we require
the existence of a \emph{dominant balance} solution in the neighborhood of the finite-time singularity, that is an exact, scale invariant  solution of the \emph{asymptotic system}
\be
\label{asym system}
\dot x=f^{\,(0)}(x),
\ee
of the form
\begin{equation} \label{dom bal}
x(\tau) =A\tau^{P},\quad \tau=t-t_*,
\end{equation}
where the function $x(\tau) =\left(\alpha\tau ^p ,\beta\tau ^q,\gamma\tau ^r,\cdots\right),$
 $A=(\alpha,\beta,\gamma,\cdots)\in\mathbb{C}^n\setminus\{{0}\},$ and $P=(p,q,r,\cdots)\in\mathbb{Q}^n.$

The second condition required for the acceptance of any given decomposition is related to the subdominant part of $f$. Suppose that we have found a decomposition where the subdominant part has weight-homogeneous components $f^{j}, j=1,\cdots , k$, that is
\be
f^{(\textrm{sub})}=f^{(\textrm{1})}+\cdots +f^{(\textrm{k})},
\ee
where each term in this expansion has components
\begin{equation}\label{sub part}
{f}_i^{\,(j)}(A\tau ^{P})=\tau ^{p_i+q^{(j)}-1}{f}_i^{\,(j)}(A),\quad i=1,\cdots, n,\,\,j=0,\cdots ,k,
\end{equation}%
for some ordered and non-negative numbers $q^{(j)}$ and $A$ in some domain $\mathcal{E}$ of $\mathbb{R}^{n}$. Intuitively this means that the given decomposition of $f$ starts with the most nonlinear part of the vector field and proceeds down to the least dominant component. From the asymptotic system (\ref{asym system}), we know that  the dominant
part, ${f}^{(0)}$, asymptotes as $\tau^{P-\textrm{diag}(1)}$, and therefore dividing both sides of Eq. (\ref{sub part}) by $\tau^{P-\textrm{diag}(1)}$ and taking the asymptotic limit to the singularity as $\tau\rightarrow 0$, we arrive at a condition that has to be satisfied if the candidate subdominant part is to be \emph{less} dominant than the dominant part of the given decomposition. This is the \emph{subdominance condition},
\be
\lim_{\tau\rightarrow
0}\frac{{f}^{\textrm{sub}}(A\t^{P})}{\tau^{P-\textrm{diag}(1)}}=0.
\ee
Any pair of the form $(A,P)$ satisfying the above conditions is called is a \emph{(dominant) balance}
of the vector field ${f}$. In general, there can be  several asymptotic splittings (decompositions) of  $f$ each one admitting several balances. Each such balance is characterized by the number of  nonzero components of the vector $A$, called the\emph{ order }of the balance, the higher order of any balance being $n$. Any balance of order lower than $n$ may correspond to a \emph{particular} solution of the system, that is a solution containing less that $n$ arbitrary constants (see Refs. \cite{split,1212.6737} for more on this terminology).

Our approach to characterizing the finite-time cosmological singularities  is an asymptotic one aiming to build formal series solutions valid near the finite-time singularity. These series solutions (one formal series for each component of the solution-vector) contain a number of arbitrary constants and can be general or particular solutions of the field equations. There is a neat way \cite{split} to determine not only the  number of the arbitrary constants but also  the exact \emph{position} of each such constant appearing in the final formal expansions. This relies on the Kowaleskaya matrix (or $\mathcal{K}$-matrix for short) of a given balance   $(A,P)$ of $f$, defined through the Jacobian of the dominant part of the vector field at the given balance,
\begin{equation}
\mathcal{K}=D{{f}}^{(0)}(A)-\textrm{diag}\,P.
\end{equation}
The $n$ eigenvalues $(\rho _{1},\cdots ,\rho _{n})$ of  of the $\mathcal{K}$-matrix are  the $\mathcal{K}$-exponents of the given balance  $(A,P)$. Using the variational equation for the dominant part of the vector field and the eigenvalue-eigenvector structure of the $\mathcal{K}$-matrix, we can Taylor estimate the asymptotic solutions and build generalized Fuchsian series expansions containing rational exponents or more general $\psi$-series solutions valid around the finite-time singularity, showing   \cite{split} that the arbitrary constants characterizing the final formal expansion will first
appear in those terms whose coefficients have indices equal to a $\mathcal{K}$-exponent.

Doing all the required steps consistently leads to an elaborate way of efficiently using vector field splittings for the study of the asymptotic properties of dynamical systems, the method of asymptotic decompositions of vector fields, cf.  \cite{split}. For a recent review of this method with references and applications to cosmological situations as well as more recent developments, the interested reader is referred to \cite{1212.6737,1301.4778}) (and references therein).

\subsection{Classification of singularities}
By a classification of  finite-time singularities we mean dealing with the problem of  finding \emph{necessary} conditions for their occurrence under general conditions on the expansion, the geometry and the matter content. Our approach to this problem is to look for such conditions as \emph{contrapositive} statements of proved sufficient conditions for geodesic completeness. In a series of works  \cite{ycb-cot02,cot04,ck05,ck07,cot07,1212.6737}, we have developed theorems of the latter category, i.e., given sufficient conditions for completeness, and proved necessary statements for singularity formation. There are three main results that give sufficient conditions for geodesic completeness, and we give a summary of these here, indicating also how the passage from completeness statements to singularity formation criteria leads to an essentially unique classification of cosmological singularities. The reader is referred to the references given above for more details and proofs of the various results.

The first theorem concerns the behaviour of the scale factor. An FRW  universe is called \emph{regularly sliced} provided the scale factor $a$ is a bounded function, $0<a(t)<\infty$. In the general case we have the following completeness result.
\begin{theorem}
Let  $(\mathcal{V},g)$ be a regularly  sliced spacetime. Then the
following are equivalent:
\begin{enumerate}
\item Each slice $(\mathcal{M}_{t},g_{t} )$ is a $g_t$-complete Riemannian manifold
\item The spacetime $(\mathcal{V},g)$ is globally hyperbolic
\end{enumerate}
\end{theorem}
According to this result, taken in the contrapositive direction,  an emerging  singularity will be either a big bang ($a\rightarrow 0$), or a big rip ($a\rightarrow\infty$). However,  as it follows from  a theorem of Ref. \cite{cot04}, slice completeness is not equivalent to \emph{space-time} geodesic completeness except when the space-time is static, we have also to allow for a third, weaker type of singularity, the sudden singularities having a complete slice, $a(t)\rightarrow a_s$. Hence, we are led to a trichotomy of spacetime singularities, which we denote as $\textrm{SF}_1$ (collapse), $\textrm{SF}_2$ (sudden), and $\textrm{SF}_3$ (big rip) type singularities.

Because the asymptotic dynamics of the universe depends also on the matter content, we have to examine the influence of the latter on geodesic completeness and singularities. To keep matters simple,  we focus here on the Friedman metrics and introduce the Bel-Robinson energy,  defined to be
\be
\mathcal{B}(t)\sim |D|^{2}+ |E|^{2},
\ee
where $|D|$ and $|E|$ are the norms of the electric parts of the Riemann double two form (the corresponding magnetic tensors are identically zero for the isotropic geometry assumed here). They read,  \be
|D|^{2}=3\left(\left(\dot{a}/{a}\right)^{2}+k/{a^{2}}\right)^{2},\quad |E|^{2}=3\left(\ddot{a}/{a}\right)^{2},
 \ee
and so  are essentially proportional to the fluid density $\rho$ and pressure $p$ respectively. We then have the following result, cf. Ref. \cite{cot07}.
\begin{theorem}\label{br}
A spatially closed Friedmann universe, expanding at a time {$t_*$},
that satisfies {$d<\mathcal{B}(t)<e$}, where $d,e$ are constants,  is causally geodesically complete.
\end{theorem}
Therefore for each one of the three singularity types $\textrm{SF}_1,\textrm{SF}_2,\textrm{SF}_3$ introduced above to monitor the asymptotic  behaviour of the scale factor, Theorem \ref{br} in its contrapositive formulation introduces four further subtypes depending on the matter asymptotes, namely,
\begin{description}
\item $\textrm{BR}_{1}:$\, $|D|< \infty \,|E|\rightarrow\infty $
\item $\textrm{BR}_{2}:$\, $|D|\rightarrow \infty,\,|E|\rightarrow\infty$
\item $\textrm{BR}_{3}:$\, $|D|< \infty ,\, |E|<\infty$
\item $\textrm{BR}_{4}:$\, $|D|\rightarrow \infty,\, |E|<\infty $
\end{description}
Hence, singularities can be classified into 12 classes according to the asymptotic  behaviours of the scale factor and the matter content of the universe on approach to a finite-time singularity at {$t_*$}. Various examples of these singularity classes are given in the references quoted above. A more recent example of an $(\textrm{SF}_2,\textrm{BR}_{3})$, that is a sudden, Tipler and Krolak weak singularity, is the generic structure found in Ref. \cite{ba-cot2013}.
We also note that the simple  classification of Ref. \cite{type} into four main types of singularities, relates with the classes introduced above as follows: Type I $\rightarrow(\textrm{SF}_3,\textrm{BR}_{2})$,  Type II $\rightarrow(\textrm{SF}_2,\textrm{BR}_{1})$, Type III $\rightarrow(\textrm{SF}_2,\textrm{BR}_{2})$, and Type IV $\rightarrow(\textrm{SF}_2,\textrm{BR}_{3})$.

The 12 singularity types introduced above can be further refined. When the universe approaches a finite-time singularity it is not only its volume ($\sim a^3$), or its matter content ($\sim\mathcal{B}(t)$) that changes, but so does the geometry of each slice as seen by an observer outside it. Thus the extrinsic curvature (proportional to the Hubble expansion rate for  the problems dealt with in this paper) also plays an important role. The following result \cite{ycb-cot02} provides the necessary connection with geodesic completeness.
\begin{theorem}
If $(\mathcal{V},g)$ is a globally hyperbolic, regularly sliced spacetime such that
for each finite {$t_{1},$  the norms of the spatial gradient of the lapse $|\nabla N|_{{g}_t}$} and of the extrinsic curvature
{$|K|_{{g}_t}$}  are integrable functions  on an interval
{$[t_{1},+\infty )$}, then $(\mathcal{V},g)$ is future causally geodesically complete.
\end{theorem}
In its contrapositive form, this theorem refines  the 12 singularity classes above by introducing three final (Hubble parameter) behaviours:
\begin{description}
\item $\textrm{H}_1$:\,  $H$ is not piecewise continuous, or
\item $\textrm{H}_2$:\, $H$ blows up in a finite time, or
\item $\textrm{H}_3$:\, $H$ is defined and integrable  for only a finite
proper time interval.
\end{description}

In this way, one may arrive at a complete, three-level classification into 36 inequivalent singularity classes of the form $(\textrm{SF},\textrm{BR}, \textrm{H})$ as determined by the asymptotic  behaviours of the scale factor (volume), Bel-Robinson energy (matter fields) and the Hubble parameter (extrinsic curvature) on approach to the finite time singularity.

\end{document}